\documentclass[review,sort&compress]{elsarticle}

\usepackage{graphicx}
\usepackage{amsmath}
\usepackage{amssymb}
\usepackage{mathrsfs}
\usepackage{subfigure}
\def\be{\begin{equation}}
\def\ee{\end{equation}}
\def\ba{\begin{array}}
\def\ea{\end{array}}
\def\bea{\begin{eqnarray}}
\def\eea{\end{eqnarray}}

\def\({\left(}
\def\){\right)}
\def\[{\left[}
\def\]{\right]}

\usepackage{lineno}
\modulolinenumbers[5]

\journal{Journal of \LaTeX\ Templates}

\begin{document}
\begin{frontmatter}

\title{Chirped nonlinear resonant states in femtosecond fiber optics}

\author[mymainaddress]{Shailza Pathania}
\author[mysecondaryaddress2]{Amit Goyal\corref{mycorrespondingauthor}}
\cortext[mycorrespondingauthor]{Corresponding author}
\ead{amit.goyal@ggdsd.ac.in}
\author[mysecondaryaddress3]{Thokala Soloman Raju}
\author[mymainaddress]{C. N. Kumar}

\address[mymainaddress]{Department of Physics, Panjab University, Chandigarh 160014, India}
\address[mysecondaryaddress2]{Department of Physics, GGDSD College, Chandigarh 160030, India}
\address[mysecondaryaddress3]{Indian Institute of Science Education and Research (IISER) Tirupati, Andhra Pradesh 517507, India}

\begin{abstract}
We show the existence of nonlinear resonant states in a higher-order nonlinear Schr\"odinger model that appertains to the wave propagation in femtosecond fiber optics, under certain parametric regime. These nonlinear resonant states are analytically illustrated in terms of Gaussian beams, Airy beams, and periodic beams that resulted due to the presence of quadratic, linear, and constant type of `smart' potentials, respectively, of the ensuing model. Interestingly, the nonlinear chirp associated with each of these novel resonant states can be efficiently controlled, by varying the self-steepening term and self-frequency shift. Furthermore, we have conducted numerical experiments corroborative of our analytical predictions.
\end{abstract}

\end{frontmatter}


\section{Introduction}
The development of all-optical soliton transmission systems is considered to be one of the hottest technologies of 21st century.
In 1973, Hasegawa and Tappert \cite{akira} theoretically predicted optical solitons in fibers. Later in 1980, Mollenaeur, Stolen and Gordon from the Bell Telephone Laboratories confirmed the existence of such phenomena experimentally \cite{gordon}.
It is a well-known fact that the optical solitons owe their existence due to the delicate balance between dispersion and self-phase modulation (SPM) \cite{akira,govind} . Owing to the prevalence of different phase sensitive nonlinear processes, only  few nonlinear effects that may arise from the nonlinear susceptibility $\chi^{(3)}$ will be present in the nonlinear fibers. Among them will be self-steepening effect and stimulated Raman scattering (SRS) effect,  because of $\chi^{(3)}$ in ultrashort pulses \cite{govind}. Additionally, these ultrashort pulses will also be subjected to group-velocity dispersion and third order dispersion (TOD). Then the model equation that describes the pulse propagation through optical waveguide  effectively, is the celebrated higher-order nonlinear Schr\"odinger
(HNLS) equation. Kodama and Hasegawa, obtained HNLS equation, for the first time, that includes higher order effects such as TOD, self-steepening, and SRS \cite{kodama1,kodama3}.
In 1985, Dianov et al. \cite{A1} discovered the phenomena of Raman-induced frequency shift.
A significant work has been done on the analysis of HNLS equation by Potasek \cite{A6} and Serkin \cite{A7}.
Grudinin et al. \cite{A8} report the first experimental study of the dynamics of the femtosecond pulses in optical fibers.
The HNLS equation has been shown to support bright / dark soliton \cite{bd2} and rogue wave solutions \cite{rw3} under different parametric conditions. Recently, dipole solitons \cite{dipole}, periodic soliton interactions \cite{liu1} and self-similar solitons \cite{harneet} has been studied in the presence of higher-order effects. For achieving ultra-high speed in long-haul telecommunication networks, solitons are transmitted at a high-pulse-repetition rate. In view of this, it is highly demanded that the higher-order effects may be retained in the HNLS equation for femtosecond pulse propagation.
\par For femtosecond pulse propagation, TOD plays pivotal role in a situation when the value of GVD is near to zero. But for pulses whose width is about 100fs, power of the order of 1 Watt, and GVD bit away from zero, the effect due to TOD can be neglected \cite{vivek}. Despite this, the significant role played by  self-steepening and self-frequency shift terms can not be simply undermined in any way, and they should be retained in the model under consideration. Thus, keeping all these effects, the HNLS model equation in dimensionless units takes the form
    \be\label{e1}
     i Q_z + \frac{1}{2} Q_{xx} +|Q|^2 Q + i \epsilon [A Q_{xxx} + B (|Q|^2 Q)_x + C Q(|Q|^2)_x] - V(x,z)Q=0.
    \ee
\par In Eq. (\ref{e1}), $Q(z,x)$ indicates the complex envelope of the electric field, $\epsilon$ indicates perturbation in which $A$ refers to TOD, $B$ refers to self-steepening and $C$ signifies self-frequency shift. And $V(z,x)$ signifies the `smart' potential \cite{soloman}.
\par We know that the cubic nonlinear Schr\"odinger equation (NLSE) or Gross-Pitaevskii eqution (GPE) with appropriate trapping potential aptly describes the dynamics of  dilute-gas Bose–Einstein condensate (BEC) \cite{Pitaevskii}. The fact that different traps used to arrest BEC has paved the way for finding new solutions of NLSE or GPE with new potentials \cite{yuri,lincoln,kevre,PRL,pk,optik,shal}. We emphasize here that, our motivation to consider these resonant states in the presence of `smart' potentials essentially, stems from this fact.
\par Now-a-days, there is renewed interest in studying the chirped pulses, as they find useful applications \cite{malomed,lisak,A2,A3}. In particular, Hmurcik and Kaup \cite{kaup} studied linearly chirped pulses with a hyperbolic-secant-amplitude profile, numerically. Following this, many authors have published their works showing the existence of chirped soliton-like solutions \cite{kruglov,A5,chen,soloman1,goyal,opt1}. Recently, the propagation of chirped optical pulses has been studied in the context of nonlinear metamaterials \cite{A10}. One of the present authors solved Eq. (\ref{e1}) in the absence of `smart' potential and obtained soliton-like solutions with nonlinear chirp \cite{pramana}. In the present work, we show the existence of nonlinear resonant states in a higher-order nonlinear Schr\"odinger model that appertains to the wave propagation in femtosecond fiber optics, under certain parametric regime. These nonlinear resonant states are analytically illustrated in terms of Gaussian beams, Airy beams, and periodic beams that resulted due to the presence of quadratic, linear, and constant type of `smart' potentials, of the ensuing model. Furthermore, we have conducted numerical experiments corroborative of our analytical predictions. In our context, these resonant states appear to be the stationary states in an optical waveguide that exhibit perfect transmission, akin to the appropriate scenario in Bose-Einstein condensates \cite{paul}.
\section{Chirped resonant states}
For the purpose of obtaining chirped resonant states, we begin our analysis by assuming solution to Eq. (\ref{e1}) in the form of
   \be\label{e2} Q(x,z) =  \rho(\xi) e^{i(\psi(\xi)-\omega z)}, \ee
where $\xi= (x-v z)$ , $v$, $\omega$  are real parameters,
$\psi(\xi)$ is the phase function and $\rho(\xi)$ is the amplitude
function. Substituting Eq. (\ref{e2}) into Eq. (\ref{e1}) and
separating the real and imaginary parts, we obtain
    \be \label{e3}\begin{aligned} v \psi^{'}\rho  & + \omega \rho + \frac{1}{2}(\rho^{''} - \psi^{'2} \rho) +  \rho^{3}
    \\&+\epsilon(-3 A
 \psi^{''}\rho^{'} - 3 A \psi^{'}\rho^{''} - A \psi^{'''} \rho + A\psi^{'3}\rho - B \rho^3\psi^{'}) - V(x,z) \rho=0,
\end{aligned} \ee
    \be\label{e4} - v \rho{'} + \rho^{'} \psi^{'} + \frac{1}{2}\psi^{''}\rho + \epsilon (A
    \rho^{'''} -3 A\psi^{'}\psi^{''}\rho -3A \psi^{'2} \rho^{'} +(2C+ B) \rho^2 \rho^{'})=0.\ee
For the femtosecond pulses, far away from zero GVD, we put $A=0$ in Eq. (\ref{e1}).
After integrating Eq. (\ref{e4}), we have an expression for $\psi^{'}$ written as
    \be\label{e5} \psi^{'} = \frac{I}{\rho^2(\xi)} + v+ \alpha \rho^2 (\xi), \ee
where $I$ is a constant of integration and  $\alpha = -\frac{\epsilon}{2}(2C+B). $
Also we put $I=0$, thus ensuring that the phase does not diverge. By substituting the expression of $\psi{'}$ into Eq. (\ref{e3}), we obtain
     \be\label{e6} \rho{''}+ (v^2+ 2\omega - 2 V(\xi))\rho +2(1-B\epsilon v)\rho^{3} -\frac{\epsilon ^2}{4}
    (2C+B)(2C-3B) \rho^5 =0. \ee
The frequency change across the pulse at any distance $z$ is known as frequency
chirp which is given by $\delta\omega(z,x)=-\frac{\partial}{\partial x}[\psi(\xi)-\omega z]=-\psi'(\xi)$.
Here, the frequency chirp depends considerably on the exact pulse shape through the relation $\delta\omega(z,x)=-\psi'(\xi)=-(v+\alpha\rho^2 (\xi))$, where $v$ and $\alpha$ are the constant and nonlinear chirp parameters, respectively.
 We note here that the parameter `$\alpha$' depends on the model coefficients---self-steepening `$B$' and self-frequency shift `$C$'. Thus, the frequency chirp can be efficiently controlled by varying these coefficients. In Eq. (\ref{e5}), for $2C=-B$, $\alpha$ comes out to be zero and corresponding solutions of Eq. (\ref{e6}) have trivial phase or also known as unchirped solutions. But, in this work, we report the chirped solutions with non-trivial phase modulation for $\alpha\neq 0$. Now, for $2C = 3B$ and $ B = \frac{1}{\epsilon v}$, Eq. (\ref{e6}) reduces to the form expressed as
    \be\label{e7} \rho^{''} - 2 V(\xi)\rho + (2 \omega + v^2 )\rho =0.\ee
For different choices of `smart' potentials $V(\xi)$---quadratic, linear and constant potential,
Eq. (\ref{e7}) yields different types of solutions such as Gaussian, Airy and periodic solutions, respectively.
Interestingly, Eq. (\ref{e7}) can be observed as quantum
mechanical Schr\"odinger equation, by identifying $V(\xi)$ as
potential and $(2 \omega_n + v^2 )$ as energy eigenvalue for corresponding $\rho_n$.
Like quantum mechanical Schr\"odinger equation, for different
value of $n$, various solutions can be generated for Eq.
(\ref{e7}).

\subsection{Chirped Gaussian beams}
In order to obtain the resonant Gaussian beams as exact solutions of this model, we take the quadratic potential, $V(\xi) = {k_1} \xi^{2}$ where ${k_1}$ is a real positive constant. In this case, Eq. (7) becomes the quantum Schr\"odinger equation for harmonic oscillator and the well-known solutions are given in terms of Hermite polynomials. Thus, we have a class of solutions $\rho_n$ for Eq. (\ref{e7}), given as
    \be \rho_n(\xi) = \sqrt{\frac{(2{k_1})^{\frac{1}{4}}}{2^n n! \pi^{\frac{1}{2}}}}~ e^{-\sqrt{\frac{{k_1}}{2}} \xi^2} ~\text{H}_n\left[(2{k_1})^{\frac{1}{4}}\xi\right], \ee
for different values of $n$ and such that
    \be \omega_n = (2n+1) \sqrt{\frac{{k_1}}{2}}-\frac{v^2}{2}.\ee

\begin{table}
\centering
\caption{Expression of $\rho_n$ and $\omega_n$ for different values of $n$\\}
    \begin{tabular}{|c|c|c|}
        \hline
        $ n $           &       $ \rho_n $      &        $\omega_n$                                                     \\ \hline
        $n=0$           & $(2{k_1})^{\frac{1}{8}} \pi^{-\frac{1}{4}}~ e^{-\sqrt{\frac{{k_1}}{2}} \xi^2} $
                                                & $  \sqrt{\frac{{k_1}}{2}}-\frac{v^2}{2}$                          \\
        $n=1$           & $\sqrt{2}(2{k_1})^{\frac{3}{8}} \pi^{-\frac{1}{4}}~\xi~ e^{-\sqrt{\frac{{k_1}}{2}} \xi^2} $
                                                & $ 3\sqrt{\frac{{k_1}}{2}}-\frac{v^2}{2}$                            \\
        $n=6$           & $\frac{(2{k_1})^{\frac{1}{8}} \pi^{-\frac{1}{4}}}{12\sqrt{5}}~ (16\sqrt{2}{k_1}^{\frac{3}{2}}\xi^6-120{k_1}\xi^4+90\sqrt{2}{k_1}^{\frac{1}{2}}\xi^2-15)~ e^{-\sqrt{\frac{{k_1}}{2}} \xi^2} $
                                                & $ 13\sqrt{\frac{{k_1}}{2}}-\frac{v^2}{2}$                           \\ \hline
    \end{tabular}
\label{table:1}
\end{table}

In Table \ref{table:1}, we have shown the expressions for
$\rho_n(\xi)$ and corresponding $\omega_n$, for $n=0,1$ and $6$, for illustrative purposes.
Thus, we get nonlinear localized Hermite modes of fixed velocity.
\begin{figure}[ht!]
\begin{center}
     \subfigure[]{\includegraphics[scale=0.8]{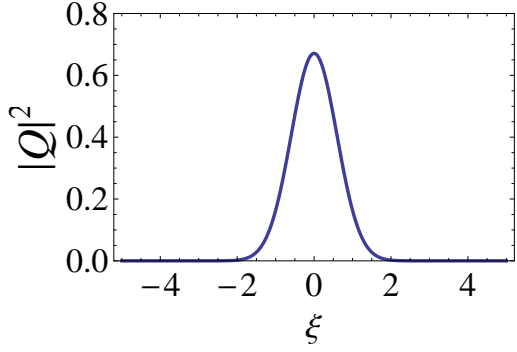}}~~~
     \subfigure[]{\includegraphics[scale=0.8]{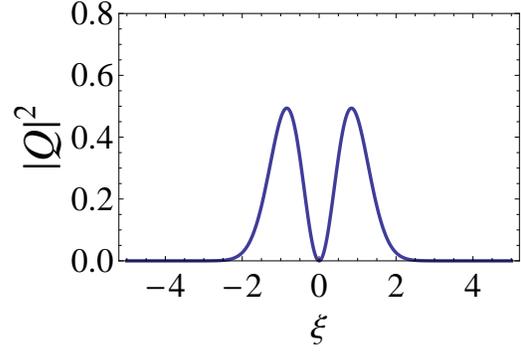}}\\
     \subfigure[]{\includegraphics[scale=0.8]{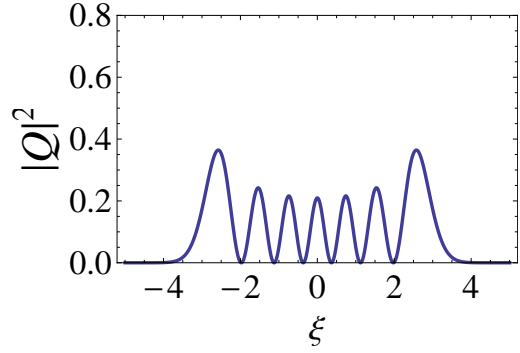}}
    \caption{\label{Fig1}Intensity profile of chirped resonant Gaussian beams for (a) $n=0$,
    (b) $n=1$ and (c) $n=6$. The other parameter used is ${k_1}=1$.}
\end{center}
\end{figure}
The complete complex wave solution for  Eq. (\ref{e1}) can be written as
    \be\label{sol1}Q_n(x,z) =  \sqrt{\frac{(2{k_1})^{\frac{1}{4}}}{2^n n! \pi^{\frac{1}{2}}}}~ e^{-\sqrt{\frac{{k_1}}{2}} \xi^2} ~\text{H}_n\left[(2{k_1})^{\frac{1}{4}}\xi\right]~e^{i(\psi(\xi)-\omega z)}, \ee
where phase profile $\psi(\xi)$ is given as
    \be\label{phase1}\psi(\xi)=v \xi -\frac{\text{erf}\left[(2{k_1})^{1/4} \xi \right]}{v}.\ee
Here erf$(x)$ is known as error function. We have depicted the intensity
distribution, $|Q(x, z)|^2$, of Gaussian beams for $n=0$, $n=1$ and $n=6$,
respectively, with ${k_1}=1$  in Fig. \ref{Fig1}. One can observe from these plots
that there is change in the intensity and the number of peaks for different modes.

\subsection{Chirped Airy beams}
\begin{figure}[ht!]
\begin{center}
     \subfigure[]{\includegraphics[scale=0.8]{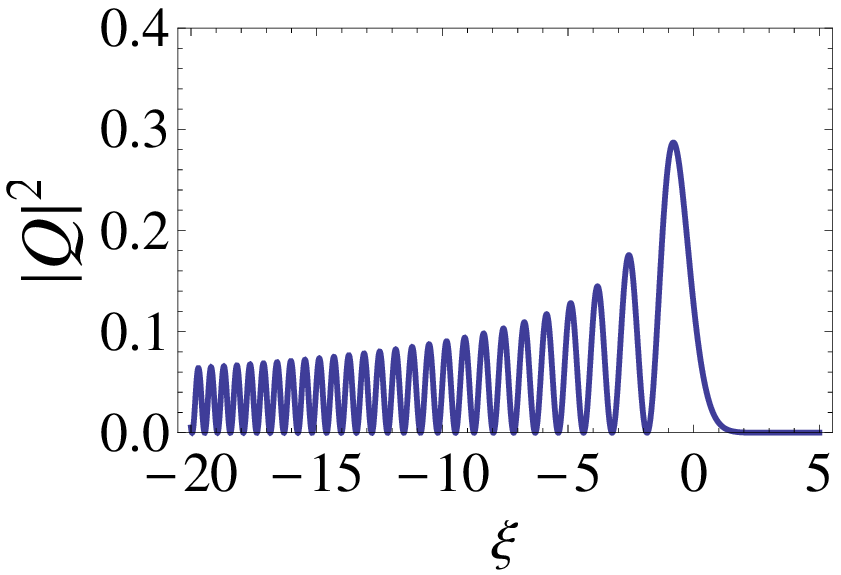}}~~~
     \subfigure[]{\includegraphics[scale=0.8]{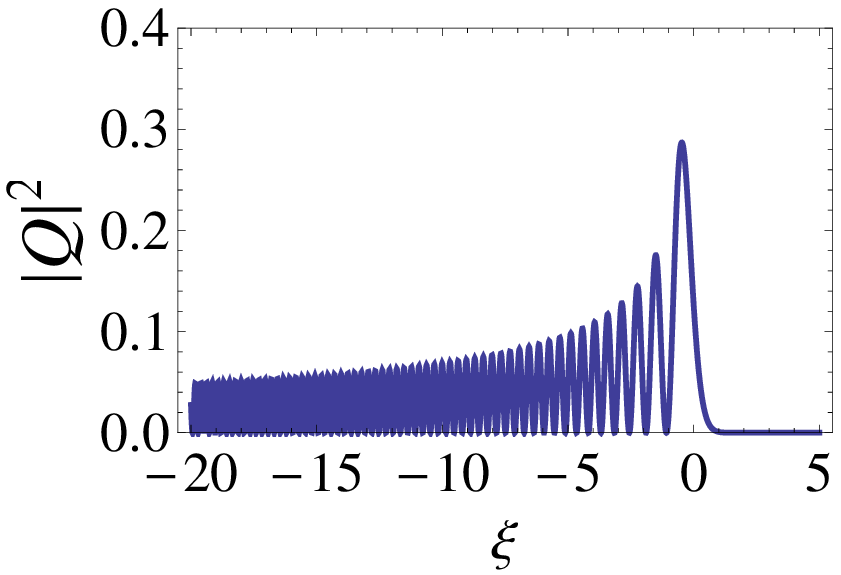}}\\
     \caption{\label{Fig2}Intensity profile of Airy beams for different
    values of parameter ${k_2}$, (a) ${k_2}=1$ and (b) ${k_2}=5$.}
\end{center}
\end{figure}

For $\omega =-\frac{v^2}{2}$, Eq.(\ref{e7}) reduces to
    \be\label{e9} \rho^{''} - 2~ V(\xi)\rho =0.\ee
For the choice of linear potential, $V(\xi) = {k_2}\xi$ where ${k_2}$ is a real positive constant, Eq. (\ref{e9}) turns out to be Airy differential equation or Stokes equation, and its solutions can be expressed as \cite{airy1}
    \be\label{e10} \rho(\xi)=\text{Ai}\left[(2~{k_2})^{\frac{1}{3}} \xi \right],\ee
where Ai$(x)$ is the Airy function. The complex wave solution for  Eq. (\ref{e1}) reads
    \be\label{sol2}Q(x,z) = \text{Ai}\left[(2~{k_2})^{\frac{1}{3}} \xi \right]~e^{i(\psi(\xi)-\omega z)}, \ee
and the phase profile is given as
    \be\label{phase2}\psi(\xi)=v \xi -\frac{2\xi}{v}\text{Ai}^2\left[(2~{k_2})^{1/3} \xi \right]+\frac{2^{2/3}}{{k_2}^{1/3} v}{\text{Ai}^\prime}^2\left[(2~{k_2})^{1/3} \xi \right],\ee
where Ai$^\prime$ denotes the derivative of the Airy function. We have depicted the intensity distribution of Airy beams for
different values of parameter ${k_2}$, in Fig. \ref{Fig2}. One can observe from these plots
that the number of resonant modes increase as we increase the value of the homogeneous parameter $k_2$ of the `smart' potential.

\subsection{Chirped periodic beams}
For constant potential $V(\xi) = {k_3}$, where ${k_3} $
is a real constant, Eq. (\ref{e7}) reduces to a homogeneous second order differential equation
    \be\label{e11} \rho^{''} + \delta\rho =0,\ee
where $\delta = 2\omega + v^2-2 {k_3}$.
Eq. (\ref{e11}) possesses periodic solution given by
    \be\label{e12} \rho(\xi)= P~\text{cos} (\sqrt{\delta} \xi )+ Q~\text{sin}(\sqrt{\delta} \xi ),\ee
where $P$, $Q$ are arbitrary constants and $\delta$ should be greater than zero which imposes a constraint condition on the wave parameter $\omega$, $\omega>k_3-\frac{v^2}{2}$.
We would like to emphasize here that the periodic solutions are exact solutions of HNLS equation, unlike the canonical free NLSE, with parametric restrictions.
\begin{figure}[ht!]
\begin{center}
     \subfigure[]{\includegraphics[scale=0.8]{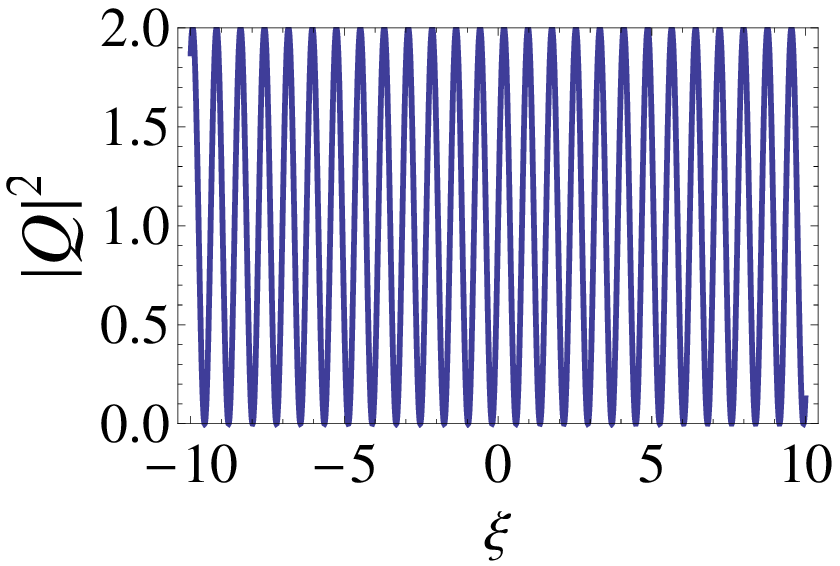}}~~~
     \subfigure[]{\includegraphics[scale=0.8]{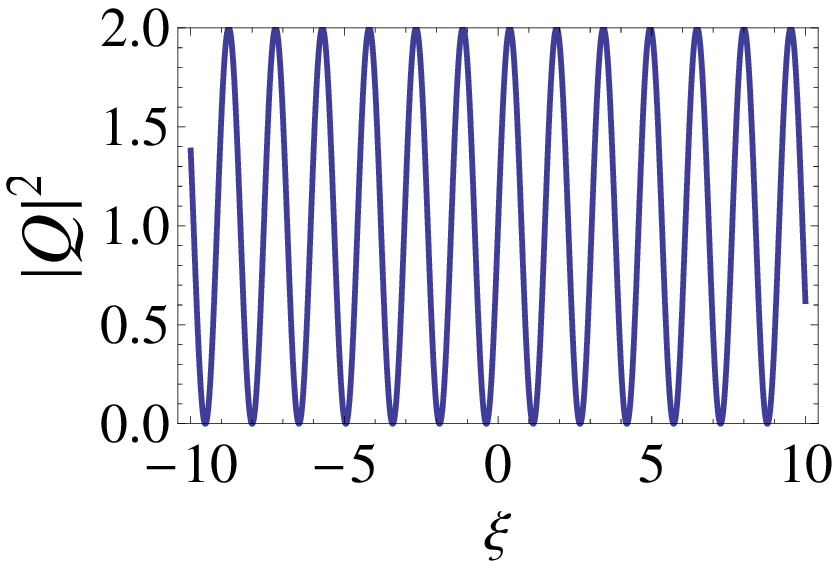}}\\
     \subfigure[]{\includegraphics[scale=0.8]{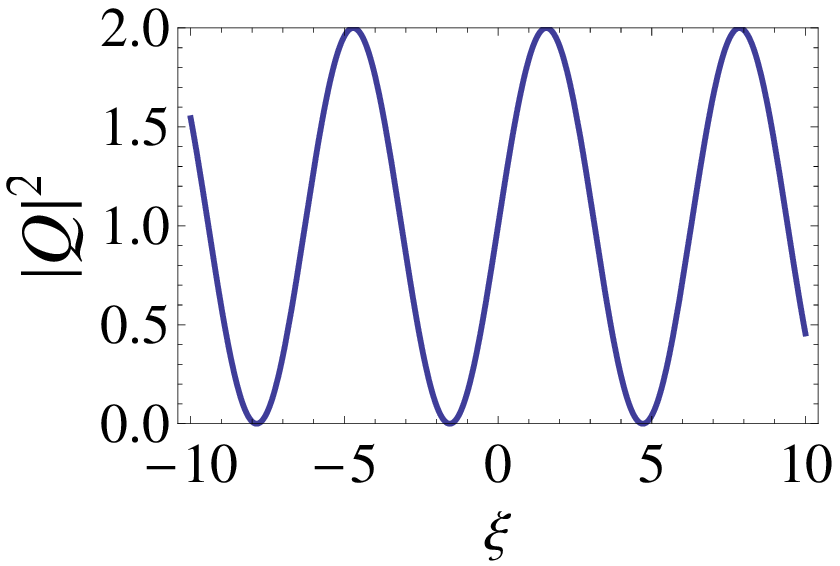}}
     \caption{\label{Fig3}Intensity profile of periodic beams for
     different values of ${k_3}$ (a) ${k_3} = 0$, (b) ${k_3} =
    6$ and (c) ${k_3} = 8$. The value of other parameters used are $P=1$, $Q=1$, $v=2.5$ and $\omega=5$.}
\end{center}
\end{figure}
The complex wave solution for  Eq. (\ref{e1}) can be written as
    \be\label{sol3}Q(x,z) = \(P~\text{cos} (\sqrt{\delta} \xi )+ Q~\text{sin}(\sqrt{\delta} \xi)\)~e^{i(\psi(\xi)-\omega z)}, \ee
and the corresponding phase profile can be obtained integrating Eq. (\ref{e5}). In Fig. \ref{Fig3}, we depicted the
intensity profile of periodic beams for different value of
parameter ${k_3}$. It is to be noted that
for ${k_3} = 0$, Eq. (\ref{e1}) returns to the standard HNLS equation which has been considered earlier \cite{pramana}. Like Airy beams,
maximum intensity of the periodic beams is also same for different
values of free parameter and it only affects the frequency of beams.

\section{Numerical simulations}
\begin{figure}[ht!]
\begin{center}
     \includegraphics[scale=0.54]{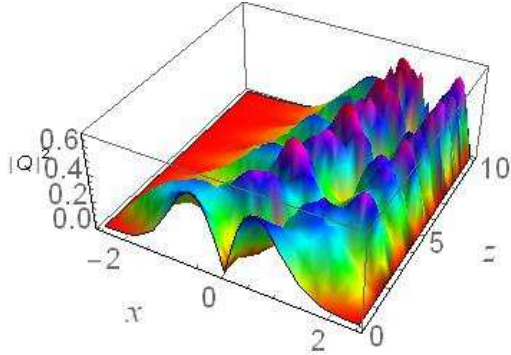}
     \caption{\label{Fig4}Plot depicting the numerical evolution of first excited resonant state for the quadratic `smart' potential.}
\end{center}
\end{figure}

In order to corroborate our analytical results with the numerical simulations, we have numerically solved Eq. (\ref{e1}), with a constraint that $2C=3B$ and in the presence of quadratic `smart' potential, using split-step Fourier method. The initial condition has been taken as the first excited resonant state. As evidenced from the surface plot (Fig. \ref{Fig4}), the numerical evolution of the first excited resonant state almost complements our analytical predictions.

\section{Conclusion}
In conclusion, we have elucidated the mechanism to generate resonant states in a higher-order NLSE that appertains  to the wave propagation in femtosecond fiber optics, under certain parametric regime. We report the existence of Gaussian beams, Airy beams, and periodic beams that resulted due to the presence of quadratic, linear, and constant type of `smart' potentials, respectively. It is observed that the free parameter in smart potentials imposes significant effects on the intensity of optical beams. Interestingly, the chirping associated with these optical beams can be efficiently controlled through self-steepening and self-frequency shift parameters. We have performed numerical simulations corroborative of the analytical results, using split-step Fourier method, and found that both analytical and numerical results are nearly complement to each other.

\section{Future work}
Recently, our group has elaborated a theoretical method, relying on isospectral deformation of Hamiltonian in supersymmetric quantum mechanics \cite{khare}, to modulate the dynamics of self-similar waves in inhomogeneous graded-index waveguide \cite{kumar,amit2}. This approach helps to construct a one-parameter dependent family of potentials and corresponding expression of wave functions for a given potential. Before the advent of supersymmetric quantum mechanics, this approach was first introduced by Infeld and Hull \cite{infeld} and Mielnik \cite{mielnik}, and was found to be useful in various physical systems \cite{cnk,dna}. Here, Eq. (\ref{e7}) can be mapped to Schr\"odinger equation of quantum mechanics which enables one to generate a class of dynamic potentials by invoking the concept of isospectral Hamiltonian approach. This way one can control the dynamical behavior of Gaussian beams. Study of dynamics of resonant states and their control using isospectral Hamiltonian approach can be well illustrated for an another interesting case of $`n(n+1)~\mbox{sech}^2{\xi}$' `smart' potential, which is analytically tractable. This work is presently under progress and will be reported elsewhere.

\section{Acknowledgment} S.P. would like to thank DST Inspire,
India, for financial support through Junior Research Fellow
[IF170725]. A.G. gratefully acknowledges Science and Engineering
Research Board (SERB), Government of India for the award of SERB Start-Up Research Grant
(Young Scientists), under the sanction no: YSS/2015/001803, during the course of this work.

\end{document}